\documentclass[fleqn,12pt,twoside]{article}
\usepackage{espcrc1}

\usepackage{graphicx}
\usepackage[figuresright]{rotating}

\newcommand{\AmS}{{\protect\the\textfont2
  A\kern-.1667em\lower.5ex\hbox{M}\kern-.125emS}}

\title{Prospects for obtaining an $r$ process from Gamma Ray Burst Disk Winds}

\author{G. C. McLaughlin\address[NCSTATE]{Department of Physics, 
        North Carolina State University, \\ 
        Raleigh, NC, 27695-8202}%
        \thanks{GCM acknowldges support from the Department of Energy under
contract DE-FG02-02ER41216}, and
        R. Surman \address[UNION]
{Department of Physics and Astronomy, Union College, \\
Schenectady, NY 12308}
 \thanks{RS acknowldges support from the Research Corporation under
contract CC5994}}

\begin{document}

\maketitle

\begin{abstract}

We discuss the possibility that $r$-process nucleosynthesis may occur 
in the winds from gamma ray burst accretion disks.  This can happen if 
the temperature of the disk is sufficiently high that
electron antineutrinos are trapped as well as neutrinos.  This implies 
accretion
disks with greater than a solar mass per second accretion rate, although 
lower accretion rates with higher black hole spin parameters may provide 
viable
environments as well.  Additionally, the outflow from the disk must either 
have 
relatively low entropy $s \sim 10$ or the initial acceleration of the wind 
must 
be slow enough that it is neutrino and antineutrino capture as opposed 
to electron
and positron capture that sets the electron fraction. 
\end{abstract}

\section{INTRODUCTION}

The $r$-process of nucleosynthesis is responsible for over half the elements 
with ${\rm A}~>~100$ but to date no self-consistent model for an astrophysical 
site for the production of these elements has 
been identified.  The two candidates most discussed in
the literature are the neutrino driven wind
of Type II supernovae  \cite{mey92,woo94}
and ejecta from neutron star mergers \cite{mey89,frei99,ros99}.  The former
site is attractive because it occurs on a timescale such that it could 
reasonably 
account
for observations in metal poor halo stars e.g. \cite{sne03} and could produce 
roughly
the right amount of material \cite{woo94}.  Furthermore, the neutrinos from
the protoneutron star set the electron fraction to a relatively low value
and this, combined with the high entropy, makes it a promising site.  Although 
attempts
to produce the $r$-process elements this way initially succeeded, subsequent 
refining of
the models indicated that the conditions were a near miss, 
e.g. \cite{hof97,fm95}. 
In addition,
self-consistent inclusion of the neutrinos in a reaction network demonstrated 
that electron neutrino capture on neutrons during the course of alpha 
particle formation 
drove the electron fraction up to values unacceptable to the $r$-process
\cite{mmf98}. 

The other primary candidate for the production of $r$-process elements,
 the ejecta from
neutron star mergers, is a viable candidate for producing some of 
these
elements. However as a dominant $r$-process site, it would be 
difficult to 
reconcile with the observations from metal poor stars, e.g. \cite{arg04}.  
If the $r$ process
has more than one component, as suggested by meteoritic evidence
\cite{was96} and observations of the low mass end of
the $r$-process distribution, then neutron star mergers could well contribute.

Here we discuss another possible site for the $r$ process, which is
in the winds from accretion disks in gamma ray bursts.  Gamma ray bursts
represent an emerging new area in nucleosynthesis research.  The
leading candidate sites for these objects are rare supernovae, 
such as a collapsars \cite{mac99}, 
and for the short bursts,
neutron star-neutron star mergers \cite{jan99,ros03}.  In either case, 
the system 
likely forms
an accretion disk surrounding a black hole.  Some nucleosynthesis will
occur in the jet \cite{pru02}, some will occur as explosive burning 
\cite{maeda03} and
still more will occur in the outflow, perhaps a wind \cite{mac99,pru04}, 
from the accretion
disk.  The rate of gamma ray bursts is estimated to be quite low,
$\sim 10^{-5}$ per year in the Galaxy, e. g. as discussed in  
\cite{zhang04}. However 
since their origin is not yet fully understood, 
they may well represent one end of a continuum of objects. 
It is therefore
difficult to estimate the rate of ejection
of their nucleosynthesis products into the interstellar medium.
\begin{figure}[htb]
\begin{minipage}[t]{8cm}
\includegraphics[angle=0,width=8cm]{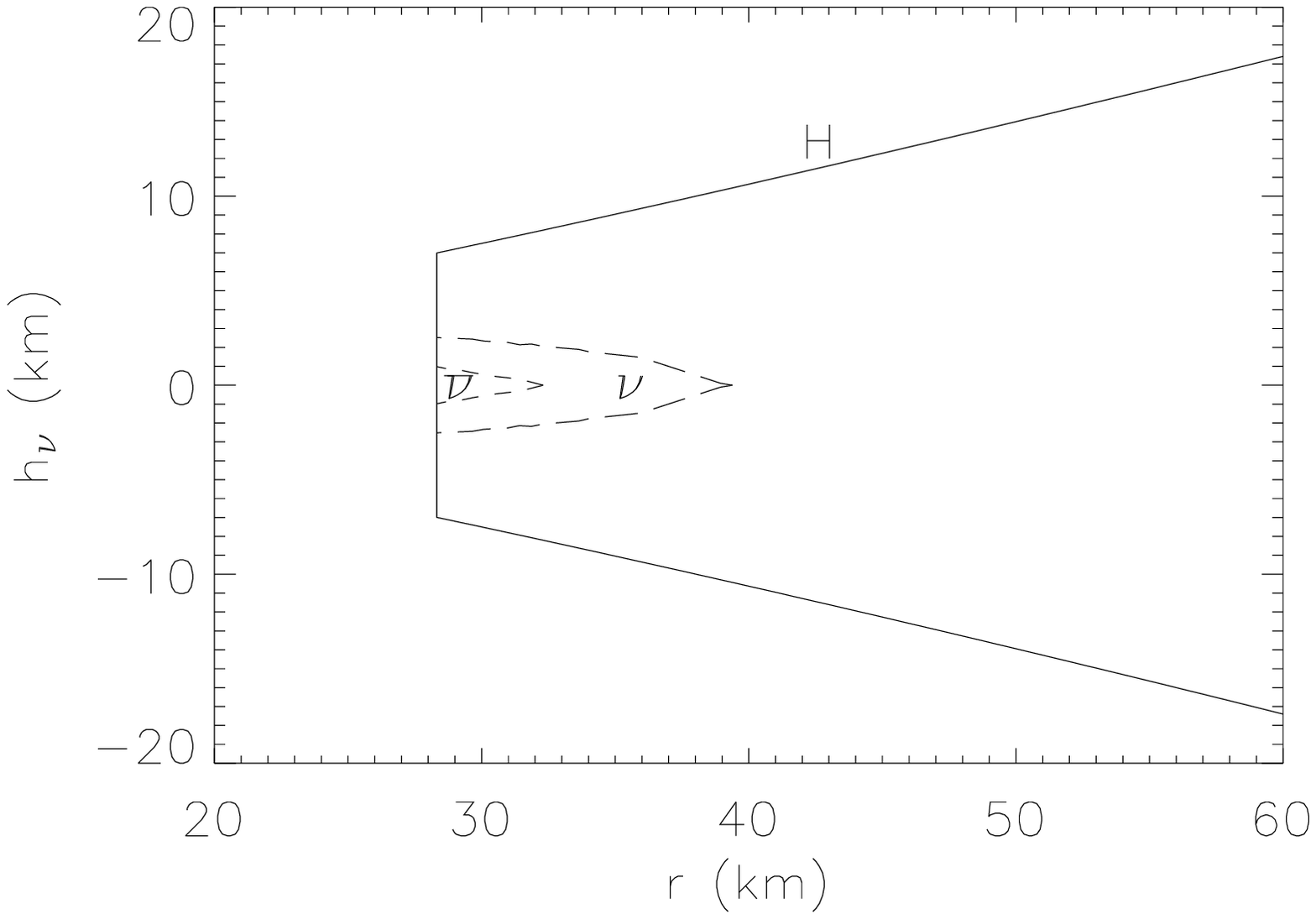}
\end{minipage}
\begin{minipage}[t]{8cm}
\includegraphics[angle=0,width=8cm]{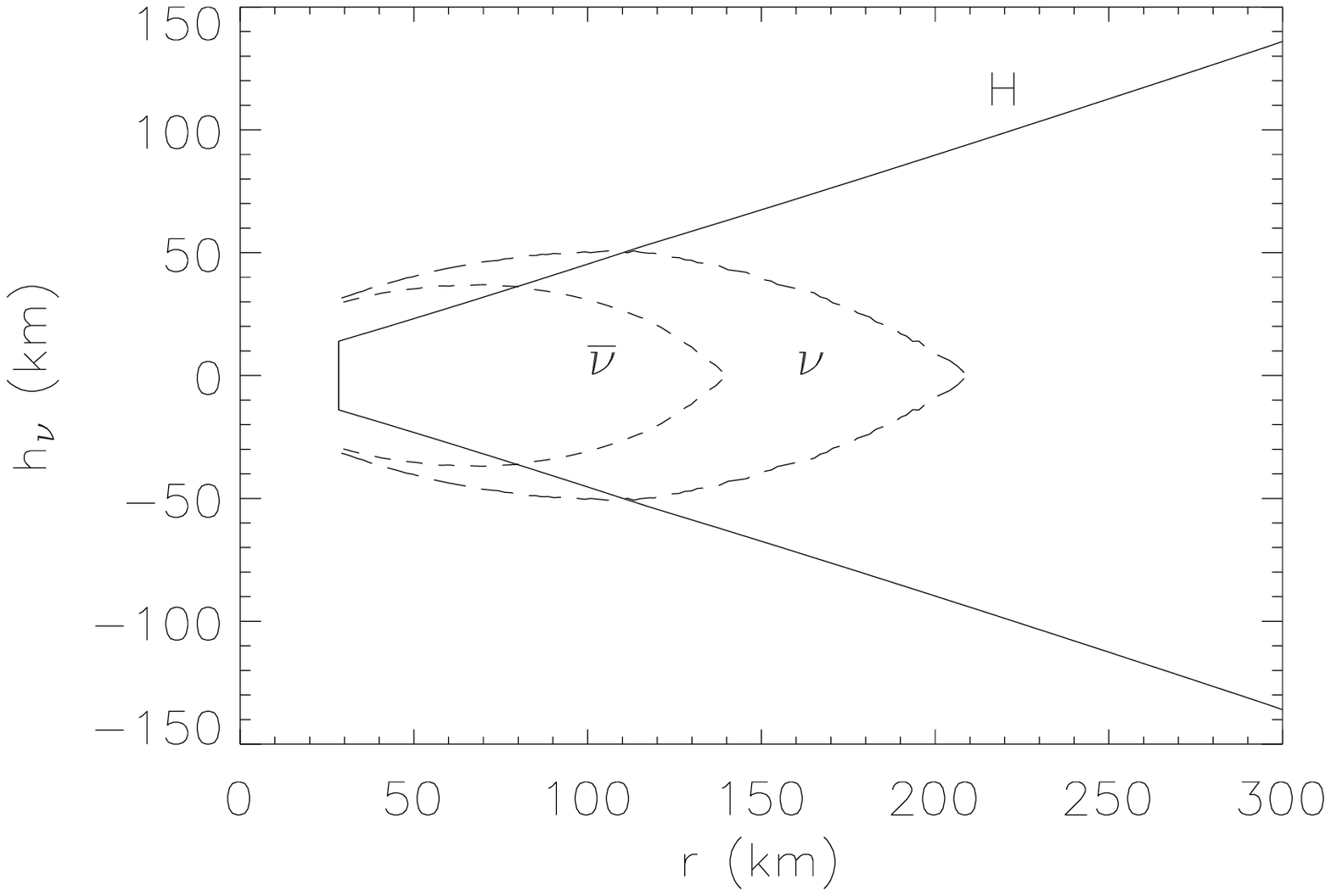}
\end{minipage}
\caption{Electron neutrino and antineutrino surfaces in  
a  $\dot{M} = 1 \, {\rm M}_\odot / \, {\rm s}$, $a=0$ (a, on the left) and
a $\dot{M} = 10 {\rm M}_\odot / \, {\rm s}$, $a=0$ (b, on the right) 
accretion disk. The solid line shows the density scale height of the
disk.
\label{fig:neutsph}}
\end{figure}

The possibility of obtaining an $r$ process in an accretion
disk wind is closely tied to 
understanding the neutrinos which are emitted copiously from the disk. 
Neutrinos are involved in all charge-changing interactions
\begin{equation}
e^- + p \leftrightarrow \nu_e + n
\label{eq:ecap}
\end{equation}
\begin{equation}
e^+ + n \leftrightarrow \bar{\nu}_e + p
\label{eq:poscap}
\end{equation}
both in the disk and in the outflow from the disk. In order to make the
outflow neutron rich, either electron capture or electron 
antineutrino capture must 
dominate.  In the case of disks with high accretion rates where both the 
neutrinos and the antineutrinos are trapped (as in Fig. \ref{fig:neutsph}), 
it is antineutrino
capture that can create conditions conducive to forming the $r$-process
elements. 

\section{NEUTRINO TRAPPING IN THE DISK}

Whether the neutrinos in the accretion disk are trapped or not
depends on the parameters that characterize the disk, such as
the accretion rate $\dot{M}$, 
the viscosity $\alpha$ and the
black hole spin parameter $a$. In Fig. \ref{fig:neutsph} we show the 
trapped regions for neutrinos
and antineutrinos which we have calculated from
 the disk models
of DiMatteo, Perna and Narayan (2002) \cite{dim02}.
\begin{figure}[htb]
\begin{minipage}[t]{8cm}
\includegraphics[angle=0,width=8cm]{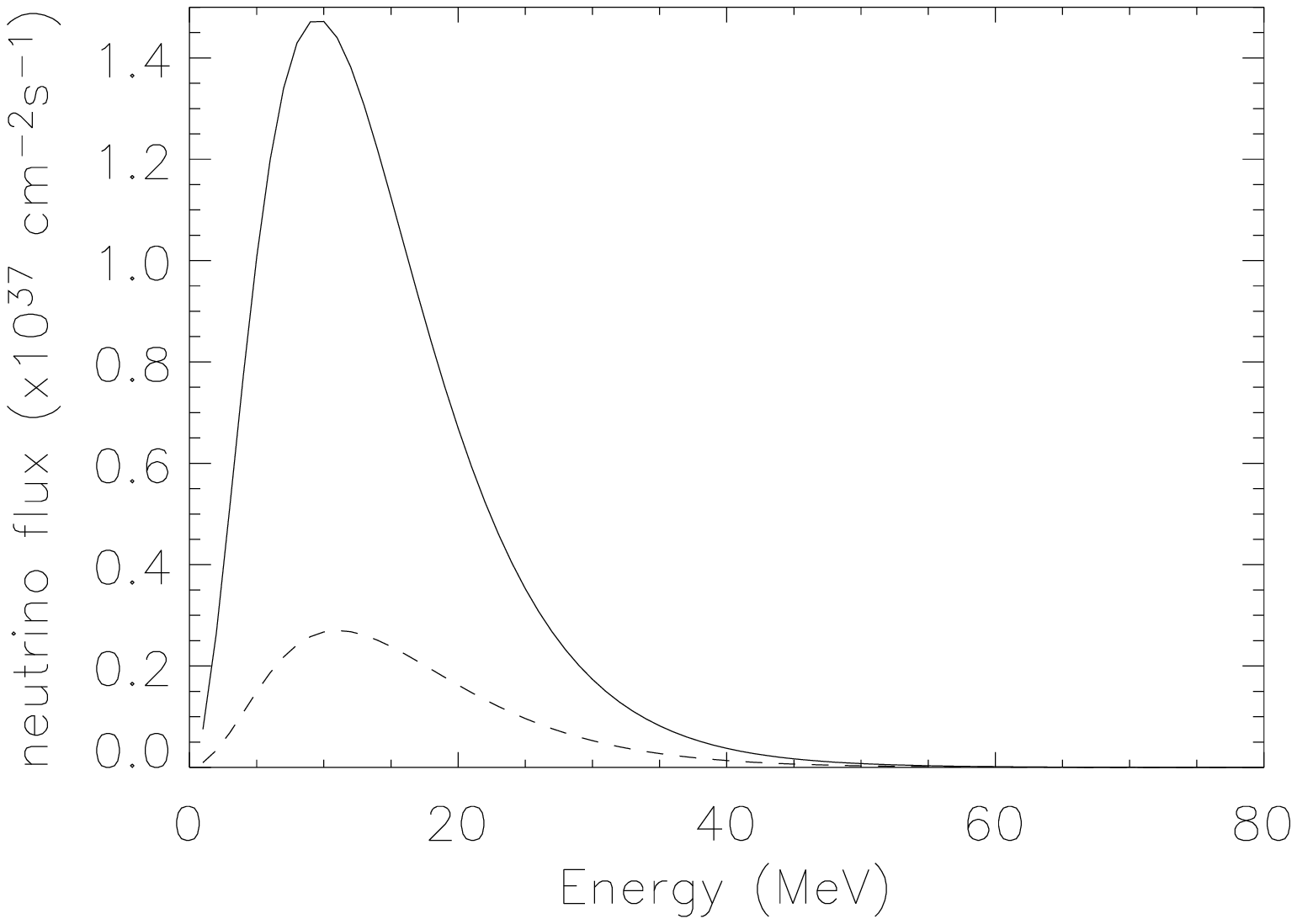}
\end{minipage}
\begin{minipage}[t]{8cm}
\includegraphics[angle=0,width=8cm]{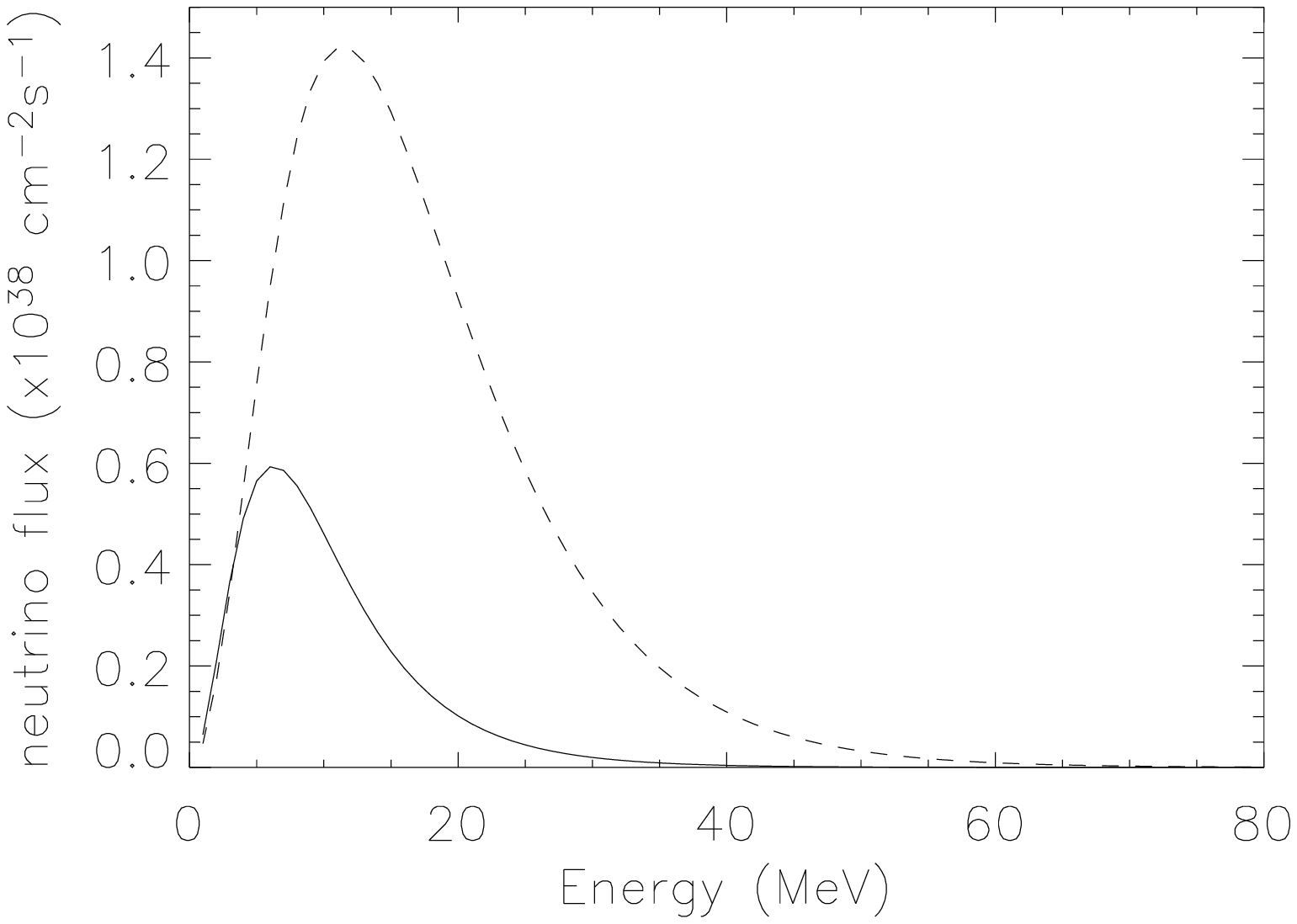}
\end{minipage}
\caption{Electron neutrino (solid line) 
and antineutrino  (dashed line) spectra at a point well above
the decoupling surfaces, $z= 220 \, {\rm km}$ and $r = 250 \, {\rm km} $
in  
a  $\dot{M} = 1 \, {\rm M}_\odot / \, {\rm s}$, $a=0$ (a, on the left) and
a $\dot{M} = 10 \, {\rm M}_\odot / \, {\rm s}$, $a=0$ (b, on the right) 
accretion disk. Note that the y-axis scale is different in the left
and right panel. 
\label{fig:spectra}}
\end{figure}
In the outer regions of the disk, neutrinos and
antineutrinos are produced by the inverse beta decay processes 
(forward reactions in Eqs. \ref{eq:ecap} and \ref{eq:poscap}) and the neutrinos
escape freely.  However, toward the center of the disk, the electron
neutrinos first become trapped, and then the electron antineutrinos become
trapped as well, enabling the backward reactions in Eqs. 
\ref{eq:ecap} and \ref{eq:poscap} to come into equilibrium with the
forward processes.
In the higher accretion rate model,  
$\dot{M} = 10 \, {\rm M}_\odot / \, {\rm s}$, $a=0$ these regions
are larger than for the more moderate accretion rate
model $\dot{M} = 1 \, {\rm M}_\odot / \, {\rm s}$, $a=0$.  In particular
the region enclosed by the antineutrino surface has expanded, although
the antineutrinos still decouple considerably further in than the neutrinos.
More detail on calculations of the disk neutrinos can be found in
\cite{sur04}.

We use the neutrino and antineutrino spectra emitted from every radius
on the disk surface to determine the neutrino and antineutrino fluxes at
every point above the disk.  The spectra from the electron neutrinos
and antineutrinos at about 200 km above the two disks shown in
Figs. \ref{fig:neutsph}a and \ref{fig:neutsph}b  can be seen
in Figs. \ref{fig:spectra}a and \ref{fig:spectra}b.  For the
lower accretion rate disk, the electron neutrino flux far dominates
the electron antineutrino flux. Although the electron antineutrinos
which come from the surface shown in \ref{fig:spectra}a have higher
temperature than the neutrinos,
the
area of the antineutrino surface is very small, so the number of 
antineutrinos is not large and a significant number of 
those seen in 
\ref{fig:spectra}a are coming directly from inverse beta decay in the 
free streaming region.  In the higher accretion rate disk,
the situation is quite different.  Because the antineutrino surface has grown,
and the antineutrinos have higher temperature than the neutrinos, the
antineutrino flux is larger and more energetic than the
neutrino flux.  In these models the electron neutrino
temperature at the neutrino surface varies with position but is  around 
$T_{MeV} \sim 2.5$ to $T_{MeV} \sim 4.5$ while the
temperature at the
antineutrino surface is around $T_{MeV} \sim 3.6$ to 
$T_{MeV} \sim 5.1$ \cite{sur04}.
In
even lower accretion rate disks (not shown here), e.g. Popham,
Woosley, Fryer (1999) \cite{pop99}, 
$\dot{M} = 0.1 {\rm M}_\odot / \, {\rm s}$, $a=0.95$, the neutrinos
are barely trapped and the antineutrinos are not trapped at all. 

\section{OUTFLOW FROM THE DISK}

We study the effect of the
charge-changing interactions on the outflow from the disk 
by using a parameterization
for the wind, as in \cite{sur04_2}. The velocity is taken to be 
\begin{equation}
|u| = v_\infty \left(1 - {R_0 \over R} \right)^\beta
\label{eq:velocity}
\end{equation}
where $R = (z^2 + r_c^2)^{0.5}$ for the first, vertical part of
the trajectory and the starting position of the 
material is $R_0$. Here $z$ is the vertical coordinate above the disk,
and $r_c$ is the cylindrically radial component along the disk.
 The parameter $\beta$ controls the acceleration of the
\begin{figure}[htb]
\begin{minipage}[t]{8cm}
\includegraphics[angle=0,width=8cm]{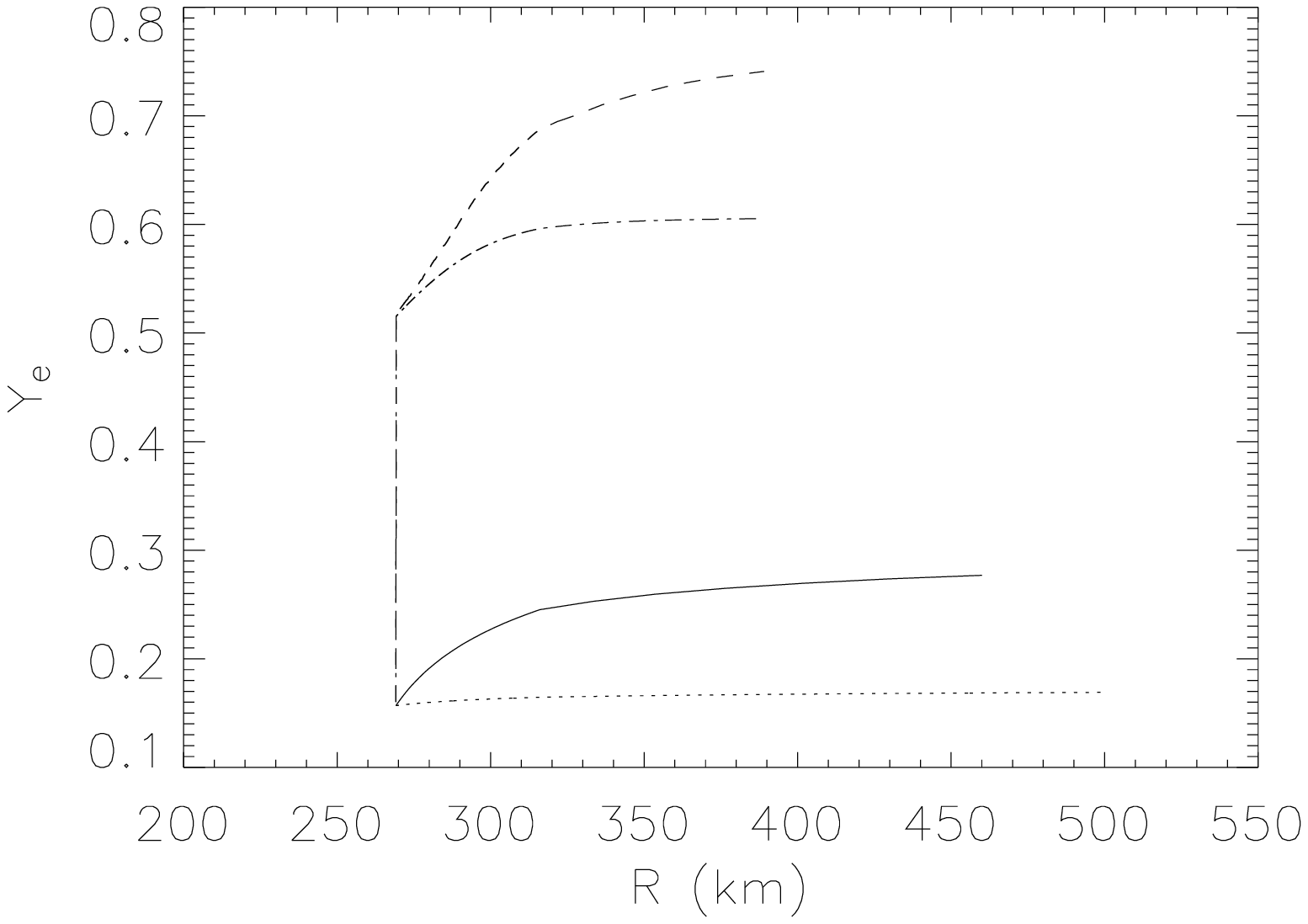}
\end{minipage}
\begin{minipage}[t]{8cm}
\includegraphics[angle=0,width=8cm]{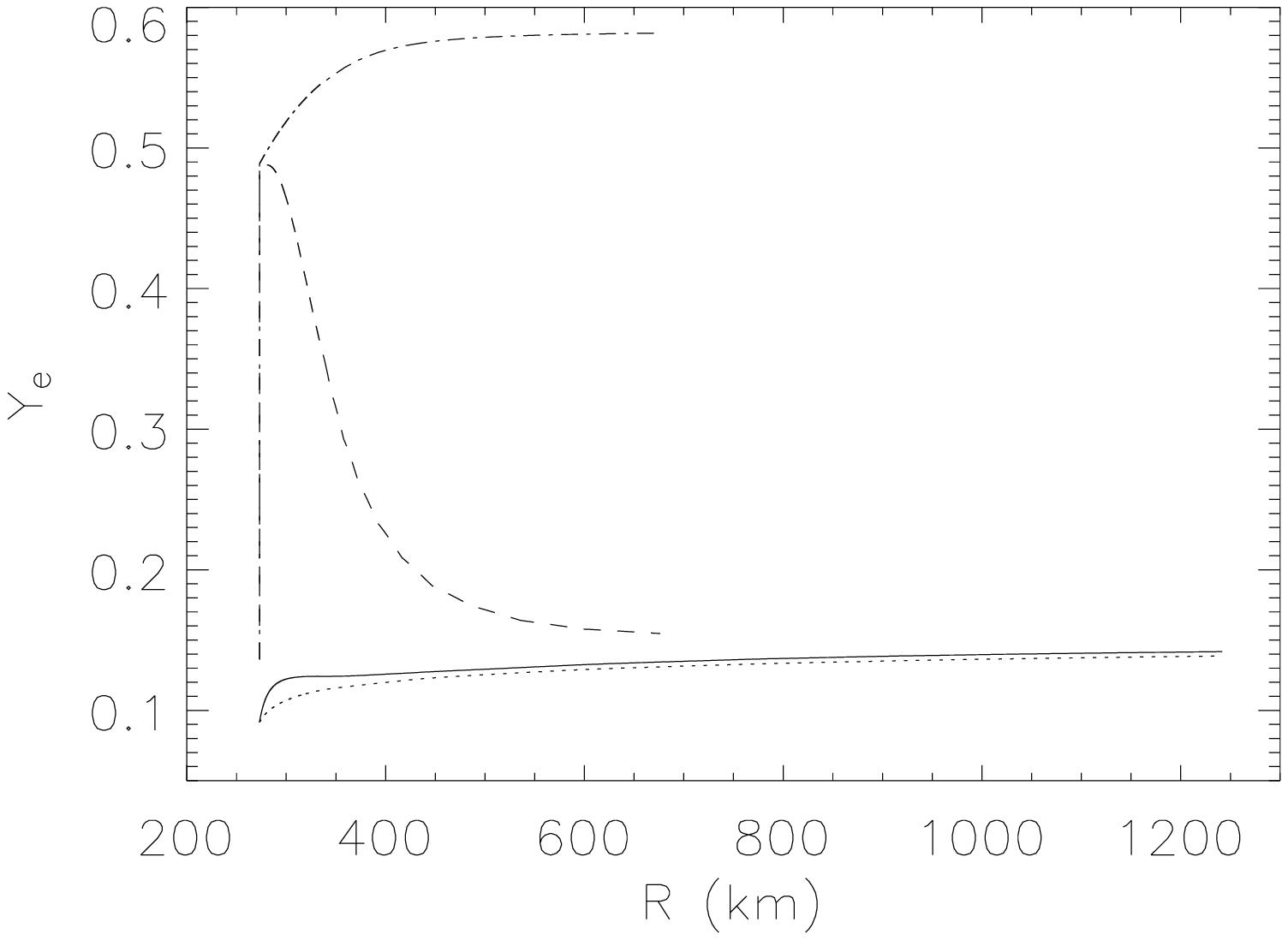}
\end{minipage}
\caption{Electron Fractions as a function of distance for two different
disk models $\dot{M} = 1 \, {\rm M}_\odot / \, {\rm s}$, $a=0$
(a, on the left) and $\dot{M} = 10 \, {\rm M}_\odot / \, {\rm s}$, $a=0$
(b, on the right) are shown.  The conditions in each case are
$s=40$, $r_0 = 250 \, {\rm km}$, $\beta = 2.5$, $v_\infty = 3
\times 10^4 \, {\rm km} \, {\rm s}^{-1}$ (dashed line) and $s=40$,
 $r_0 = 250 \, {\rm km}$, $\beta = 0.8$, $v_\infty = 3 
\times 10^4 \, {\rm km} \, {\rm s}^{-1}$ (solid line). The dotted
and dot-dashed lines show the same calculation without the neutrinos
included.
\label{fig:yevsr}}
\end{figure}
wind; for lower $\beta$ the wind accelerates faster.  In terms of the
neutrinos, larger $\beta$ means more time for the neutrinos to influence 
the composition of the outflow.  In Fig \ref{fig:yevsr}, we show outflow from
two different types of disk models, DPN 
$\dot{M} = 10 \, {\rm M}_\odot / \, {\rm s}$ and 
$\dot{M} = 1 \, {\rm M}_\odot / \, {\rm s}$. 

It can be seen from these figures that the neutrinos influence the
electron fraction in opposite directions in the two models.  In
the more moderate accretion rate model 
$\dot{M} = 1\, {\rm M}_\odot / \, {\rm s}$, $a=0$, the large flux of
neutrinos shown in Fig \ref{fig:spectra} raises the electron
fraction to quite high values. However in the
higher accretion rate model,
$\dot{M} = 10 \, {\rm M}_\odot / \, {\rm s}$, $a=0$, the large
flux of antineutrinos decreases the electron fraction to quite low values,
even in the case of a high entropy outflow.
Also interesting is the low entropy case in Fig. \ref{fig:yevsr}b,
where the electron fraction remains quite low regardless of the influence
of the neutrinos. This is due to the electron degeneracy of the material at
a high density and low entropy.

\begin{figure}[htb]
\begin{minipage}[t]{8cm}
\includegraphics[angle=0,width=8cm]{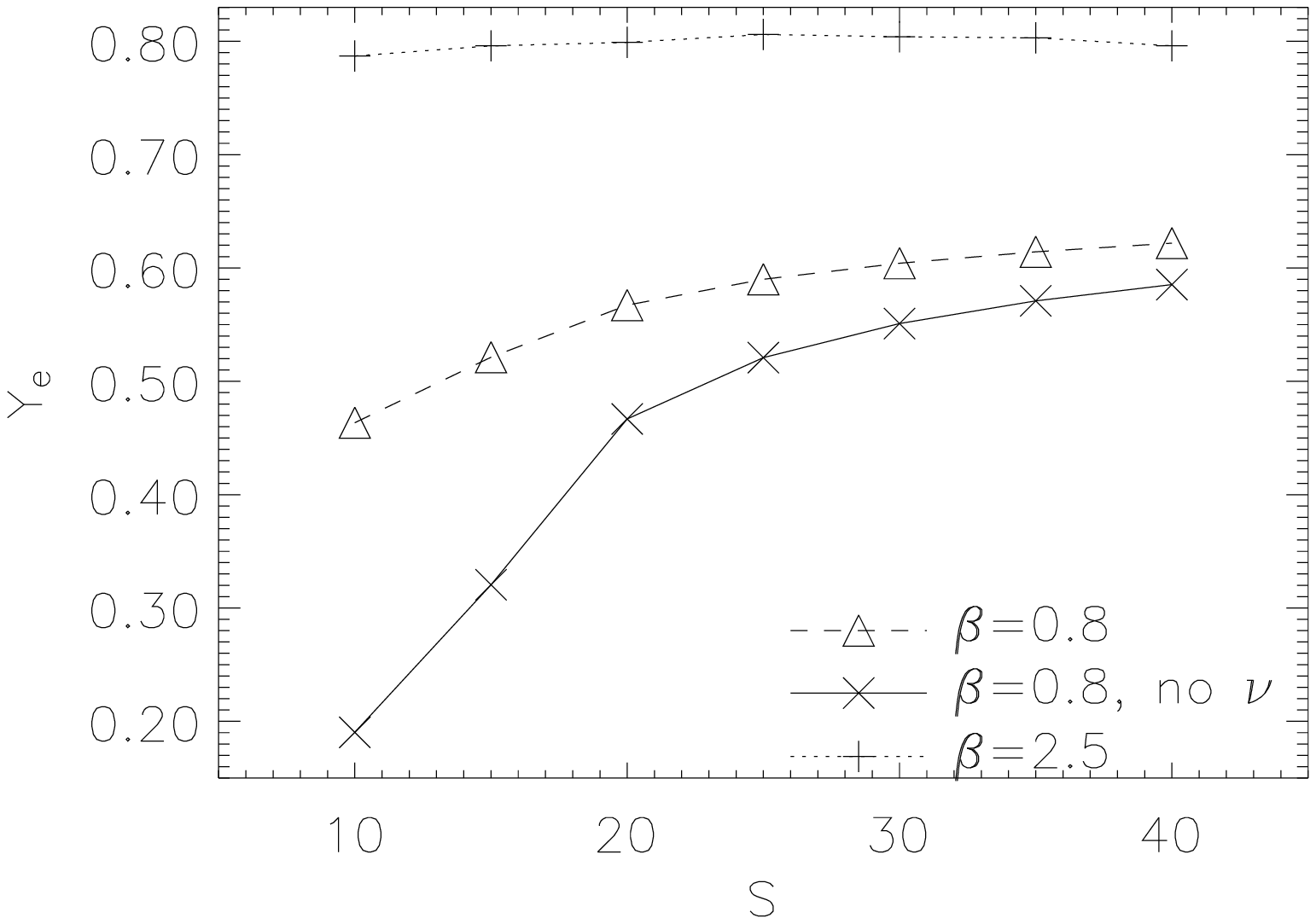}
\end{minipage}
\begin{minipage}[t]{8cm}
\includegraphics[angle=0,width=8cm]{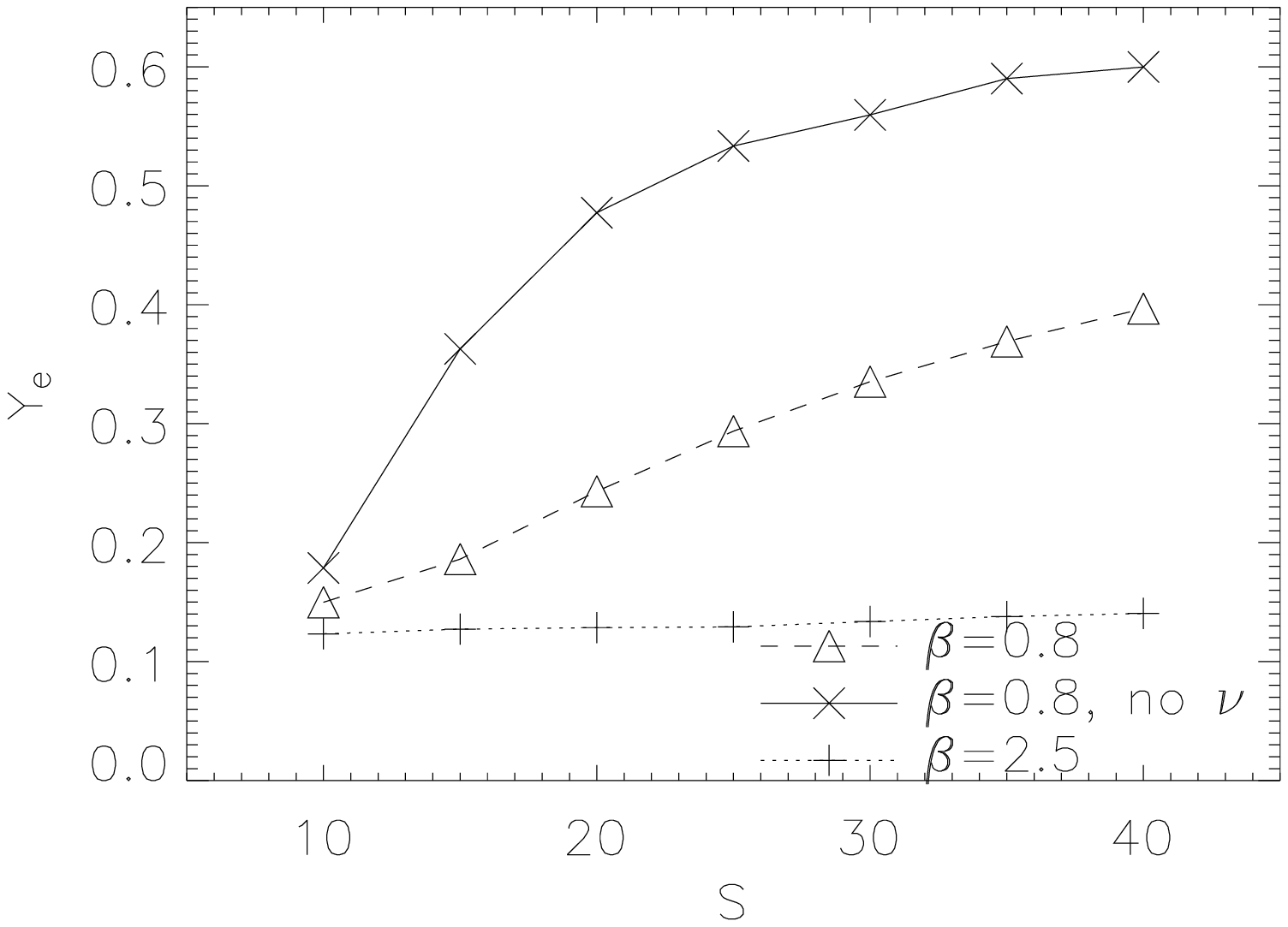}
\end{minipage}
\caption{Electron fractions in the outflow measured at a temperature of 
1 MeV for two different
disk models $\dot{M} = 1 {\rm M}_\odot / \, {\rm s}$, $a=0$
(a, on the left) and $\dot{M} = 10 {\rm M}_\odot / \, {\rm s}$, $a=0$
(b, on the right) are shown.  Various values of the 
acceleration parameter
$\beta$ are plotted against entropy per baryon in the outflow.
\label{fig:outflow}}
\end{figure}

In Fig. \ref{fig:outflow}, we survey a number of different outflows,
by plotting the electron fraction measured at a little under an MeV
against entropy for several different values of $\beta$.  In all cases
the final outflow velocity is
$v_\infty = 10^4 {\rm km} \, {\rm s}^{-1}$.  

The plus signs show the effect of a slow outflow parameter $\beta = 2.5$.  In
this case the neutrino and antineutrino capture rates completely 
overwhelm the electron and positron capture rates, and the system finds
a weak equilibrium such that 
\begin{equation}
Y_e = {1 \over 1 + {\lambda_{\bar{\nu}_ep} \over \lambda_{\nu_en}}}
\end{equation}
where $\lambda_{\bar{\nu}_ep}$ is antineutrino capture on protons and 
$\lambda_{\nu_en}$ is neutrino capture on neutrons.

Again we see that a slow outflow from a high accretion rate disk will
produce a low electron fraction due to the neutrino interactions,
although for any outflow a low entropy will also create a low electron
fraction.  For the lower accretion rate disk and slow outflows,
the neutrinos always make the material more proton rich and in
fact it is only the case of low entropy and minimal neutrino 
interactions that produces a low electron fraction.

\section{NUCLEOSYNTHESIS}

\begin{figure}[htb]
\includegraphics[angle=0,width=16cm]{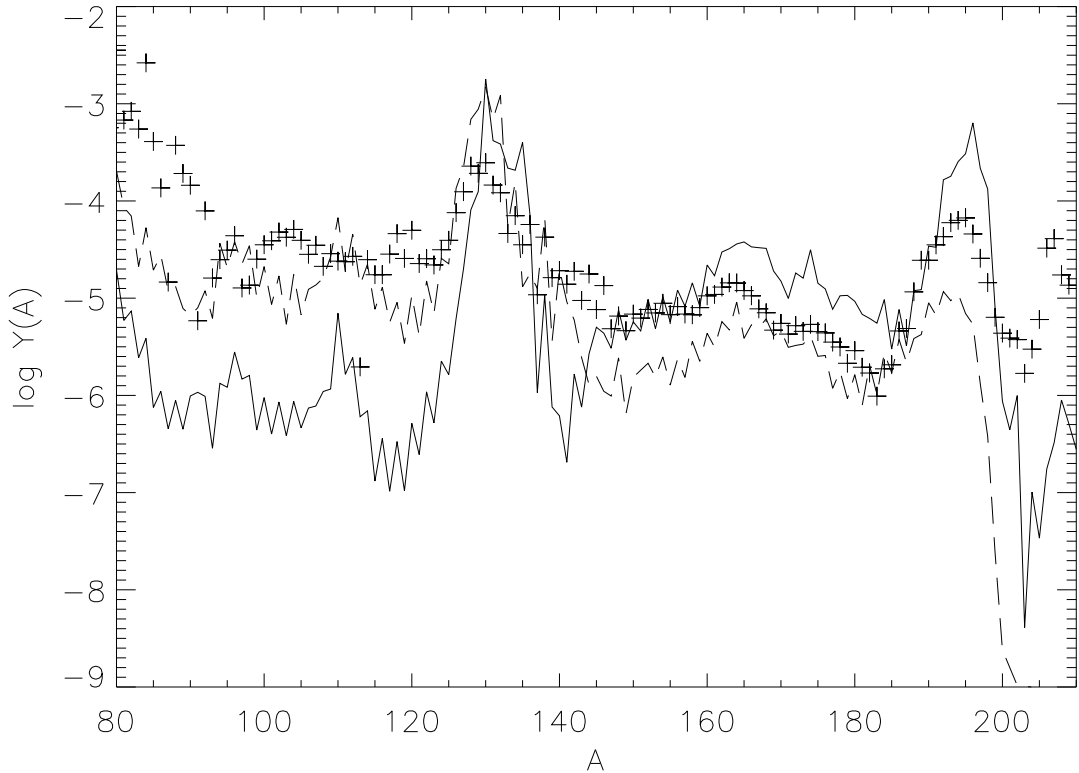}
\caption{Shows two $r$-process nucleosynthesis calculations, both in outflows
from a $\dot{M} = 10 \, {\rm M}_\odot / \, {\rm s}$, $a=0$ accretion
disk. We show a low entropy  $s=10$ outflow
with $\beta = 0.8$ and $v_\infty = 3 \times 10^4 \, {\rm km} \, {\rm s}^{-1}$
as the solid line and
 we show a high entropy outflow, $s=40$ but with a slow acceleration
parameter, $\beta = 2.5$ and  $v_\infty = 3 \times 10^4 \, {\rm km} \, {\rm s}^{-1}$ as the dashed line. 
The crosses show the measured 
solar abundances. 
\label{fig:rprocess}}
\end{figure}

Although there are two regions of parameter space which will produce
very neutron rich winds, this is not a guarantee of a successful $r$ process,
so in this section we investigate this possibility.    
We take the high accretion
rate model $\dot{M} = 10 \, {\rm M}_\odot / \, {\rm s}$, $a=0$ and an outflow
with a fairly low entropy, s=10 and $\beta = 0.8$  and a final velocity of 
$3 \times 10^4 {\rm km} \, {\rm s}^{-1}$. This is the same trajectory as shown 
as the solid line in Fig. \ref{fig:yevsr}b. Our calculation indicates that
the low entropy trajectories do produce an $r$ process as shown in 
the solid line in
Fig. \ref{fig:rprocess}.  
The dashed line shows
 a calculation with a higher entropy, $s=40$, 
but a slower outflow acceleration $\beta = 2.5$ so that the neutrinos
have greater influence.  Note that in neither of these calculations
is there an ``alpha effect'' \cite{fm95,mfw96} and this is due to
the high outflow velocity at the time alpha particles are forming.

While the higher accretion rate disks may be a viable site for the $r$ process,
moderate accretion rates disks such as  $\dot{M} = 1 \, {\rm M}_\odot / \, {\rm s}$, $a=0$
will produce  much higher electron fractions.  In
Fig. \ref{fig:outflow}a, it can be seen that the material which has been irradiated by 
a large neutrino fluence  because of its slow acceleration ($\beta = 2.5$) 
is very proton rich.
This type of outflow will produce primarily nickel, but it will also produce some
nuclei on the proton rich side of the valley of beta stability as well, such as
$^{58}{\rm Cu}$, $^{59} {\rm Zn}$,$^{50}{\rm Fe}$
and $^{52}{\rm Fe}$.  For still lower accretion rate models, such as 
$\dot{M} = 0.1 \, {\rm M}_\odot / \, {\rm s}$, $a=0.95$
the outflow will
have an electron fraction which is closer to $Y_e = 0.5$, although the neutrinos
may drive the electron fraction up to as much as $Y_e = 0.6$.  Winds of
this type with entropies of order $s \sim 30$ may have an unusual
nucleosynthesis pattern with large overproduction factors of elements such as
$^{42}{\rm Ca}$ and
$^{45}{\rm Sc}$, $^{46}{\rm Ti}$, $^{49}{\rm Ti}$, $^{63}{\rm Cu}$,
$^{64}{\rm Zn}$ as discussed in \cite{psm04}.  Lower spin parameter models
are discussed in \cite{fuj04}.

\section{CONCLUSIONS}

The winds from accretion
disks surrounding black holes in the context of gamma ray bursts
are a new arena in which to investigate nucleosynthesis.  The
most important parameters which determine the elements formed in the
outflow from the disk are the accretion rate and black hole spin parameter.
This is because the density and temperature of the disk determine where
the neutrinos and antineutrinos become trapped. Neutrinos become trapped
before antineutrinos, and in some disks
only neutrinos are trapped, not antineutrinos.  In these disks the wind
is proton rich, from the reaction $\nu_e + n \rightarrow e^- + p$,
and produces considerable nickel-56, but also elements on
the proton rich side of the valley of beta stability.
However
as the accretion rate increases, a sizable region of trapped antineutrinos
can develop, and these antineutrinos cause the wind to 
become neutron rich.
This happens because the higher temperature antineutrinos  
cause $\bar{\nu}_e + p \rightarrow e^+ +n$
to be faster than the corresponding reaction for neutrinos.  
If the outflow is
is slow enough, for example in our parameterization 
at around 
 $\beta = 2.5$, or has low entropy $s \sim 10$, then
the electron fraction is sufficiently low that $r$-process 
elements are produced in the outflow.

\end{document}